# Magnetic reversal and pinning in a perpendicular zero moment half-metal

**Authors**

N. Teichert[1], G. Atcheson[1], K. Siewierska[1], M. N. Sanz-Ortiz[2], M. Venkatesan[1], K. Rode[1], S. Felton[2], P. Stamenov[1], J. M. D. Coey[1]

**Affiliations**

1 CRANN, AMBER and School of Physics, Trinity College Dublin, Dublin 2, Ireland

2 Centre for Nanostructured Media (CNM), School of Mathematics and Physics, Queen's University Belfast, Belfast, BT7 1NN, UK

## Abstract

Compensated ferrimagnets are promising materials for fast spintronic applications based on domain wall motion as they combine the favourable properties of ferromagnets and antiferromagnets. They inherit from antiferromagnets immunity to external fields, fast spin dynamics and rapid domain wall motion. From ferromagnets they inherit straightforward ways to read out the magnetic state, especially in compensated half metals, where electrons flow in only one spin channel. Here, we investigate domain structure in compensated half-metallic $Mn_2Ru_{0.5}Ga$ films and assess their potential in domain wall motion-based spin-electronic devices. Our focus is on understanding and reducing domain wall pinning in unpatterned epitaxial thin films. Two modes of magnetic reversal, driven by nucleation or domain wall motion, are identified for different thin film deposition temperatures ($T_{\text{dep}}$). The magnetic aftereffect is analysed to extract activation volumes ($V^*$), activation energies ($E_A$), and their variation ($\Delta E_A$). The latter is decisive for the magnetic reversal regime, where domain wall motion dominated reversal (weak pinning) is found for $\Delta E_A < 0.2$ eV and nucleation dominated reversal (strong pinning) for $\Delta E_A > 0.5$ eV. A minimum $\Delta E_A = 28$ meV is found for $T_{\text{dep}} = 290°C$. Prominent pinning sites are visualized by analysing virgin domain patterns after thermal demagnetization. In the sample investigated they have spacings of order 300 nm, which gives an upper limit of the track-width of spin-torque domain-wall motion-based devices.

## Introduction

Zero-moment half-metals (ZMHM) have properties that could make them ideal materials for domain-wall motion-based spintronic devices. However, an understanding of domain formation and pinning in these materials is a precondition for controlled and reproducible application.

Zero net magnetization in ZMHMs is achieved by compensation of two inequivalent antiparallel magnetic sub-lattices made up of atoms occupying different lattice sites. Since, in general, the temperature-dependence of each magnetic sublattice is different, full compensation is achievable only at the magnetization compensation temperature $T_{\text{comp}}$.

The half-metallicity is caused by a spin gap that persists even at zero net magnetization and allows readout of the magnetic state by common spintronic measurements like anomalous Hall effect (AHE) or tunnel magnetoresistance (TMR), even at $T_{\text{comp}}$ when there is no net magnetization, This material class was predicted in 1995 by van Leuken and de Groot [1] and experimentally realized in 2013 [2] with the near-cubic films of a Heusler alloy $Mn_2Ru_xGa$ (MRG), which exhibits a combination of negligible saturation magnetization ($M_s$), high spin polarization, and high magnetic ordering temperature $T_C$. $T_{\text{comp}}$ can be tuned by variation of the Ru content in $Mn_2Ru_xGa$ (0.4 < x < 0.9). The compound crystallizes in the inverse Heusler (XA) structure with two opposing magnetic sublattices on the $4a$ and $4c$ sites, that are both occupied by Mn atoms. However only the $4c$ sublattice contributes significantly to the electron density at the fermi level. MRG is a cubic material which acquires perpendicular anisotropy in thin films by a ~1% tetragonal distortion of the unit cell due to biaxial epitaxial strain when grown on MgO substrates.

We have previously shown that the magnetic domains in this zero moment material can be visualized directly using Kerr microscopy due to a sizable magneto-optical Kerr effect (MOKE) [3] arising mainly from the $4c$ sublattice [4]. Related to this, MRG exhibits an exceptionally large anomalous Hall effect [5] allowing electrical read-out of the magnetic state. In a potential device, magnetic domains can be nucleated either electrically via spin-orbit-torque switching [6] or optically via single pulse all-optical switching [7]. Field- and current-induced domain wall motion has been shown to exhibit maximum mobility and non-saturating domain wall velocity at the angular momentum compensation point (which may differ from the magnetic compensation point $T_{\text{comp}}$) for the compensated amorphous ferrimagnets GdCoFe [8] and TbCo [9]. Since the coercivity diverges at $T_{\text{comp}}$ the domain images may never be truly characteristic of the zero-moment state because one must approach it from above or below via states with a small net moment. Evolution of the domain structure will be arrested when the coercivity exceeds the perpendicular magnetization.

Here we compare the magnetic domain patterns and time dependence of the magnetic reversal process depending on the tetragonal distortion of the material. This was modified by variation of the substrate temperature during deposition while maintaining composition, film thickness and surface roughness. Increased substrate temperatures lead to relaxation of the material, reducing the degree of tetragonal distortion, and altering the variance of the local anisotropy.

Depending on the deposition temperature we observe two distinct magnetization reversal processes, one dominated by domain wall motion and the other by domain nucleation, similar to those observed in other perpendicular thin films. [10,11] We analyse them by time-dependent measurements of the magnetic aftereffect, yielding estimations of the activation volumes involved and the activation energy distributions, whereby a strong dependence on the deposition temperature is established. The distribution of activation energy is the underlying cause of the reversal process and itself is governed by crystalline defects such as misfit dislocations that locally alter the distortion and thereby the anisotropy of the material. The defects act as pinning centres for the magnetic domains.

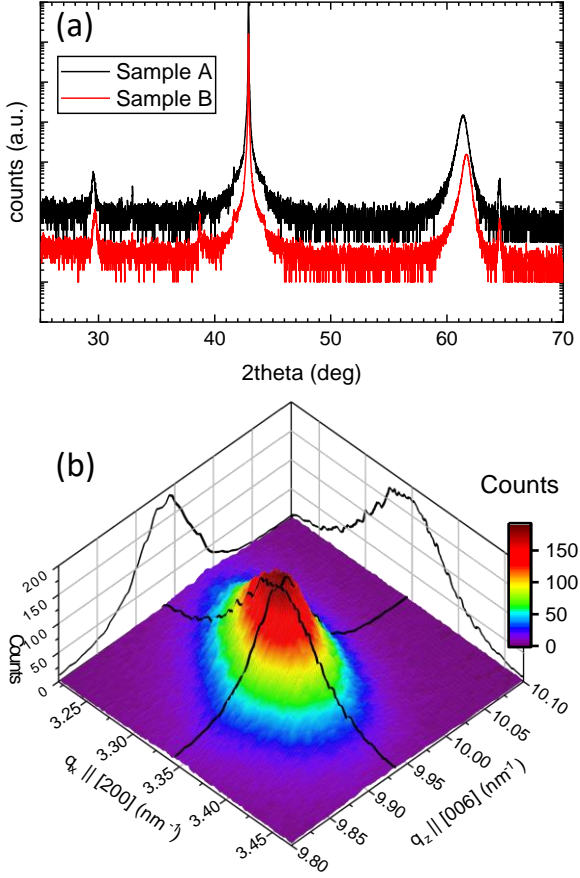

Figure 1 Structural characterization of samples A and B. (a) X-Ray diffraction patterns showing the (002) and (004) peaks of MRG due to epitaxial growth. (b) Reciprocal space maps of the MRG (206) reflection of sample B. The 3d plot is the full peak in the $q_x$-$q_z$ plane and the black curves and their projections are cross sections in $q_x$ and $q_z$, respectively, through the maximum.

The second part of this paper is focused on virgin domain patterns in MRG films. We explore the temperature dependence of the domain pattern and find that it does not change once it nucleates just below $T_C$. We also demonstrate a method for visualization of the pinning centres, based on analysis of virgin domain patterns upon repeated thermal demagnetization.

# Methodology

Epitaxial thin films of MRG were grown by DC magnetron co-sputtering on 10x10 mm$^2$ MgO(100) substrates using a Shamrock sputtering system. The films were deposited from three 75 mm targets of Mn$_2$Ga, Ru and Mn$_3$Ga in an argon atmosphere. The substrate temperature during deposition, $T_{dep}$, was set by a heating coil behind the substrate, which, unless stated otherwise, was backcoated with tantalum for improved infrared absorption. The samples A and B compared in this paper have the same composition of Mn$_{2.2}$Ru$_{0.5}$Ga and were prepared with $T_{dep} = 300°C$ (sample A) and 320°C (sample B). Sample C has a composition of Mn$_{2.2}$Ru$_{0.9}$Ga and was nominally deposited at 340°C, but without tantalum back coating. The films have a thickness of 52 nm and were capped in-situ with a 2 nm layer of aluminium deposited at room temperature to prevent oxidation. A Bruker D8 Discovery X-ray diffractometer with a copper tube (Cu K$_\alpha$ $\lambda$ = 1.59056 Å) and a double-bounce Ge[220] monochromator was used to determine the reciprocal space maps of the thin films. Other diffraction patterns and low-angle X-ray reflectivity (used to confirm the film thickness) was measured with a Panalytical X'Pert Pro diffractometer using Cu K$_\alpha$ radiation.

Domain imaging, magnetic hysteresis and aftereffect measurements were performed in a perpendicular-field electromagnet using an Evico polarization microscope illuminated by either red (central wavelength $\lambda$ = 632 nm) or blue ($\lambda$ = 455 nm) light from an LED array. For high-resolution imaging at room temperature, a 100x immersion lens (NA=1.3) and blue light were used. Lower resolution imaging and magnetic hysteresis and aftereffect measurements were done using red light and 20x magnification. For magnetic hysteresis measurements the polar Faraday effect was compensated using the mirror technique described in [12]. For temperature-dependent measurements the films were covered by a protective 10 nm film of SiO$_2$. A 50x non-immersion lens (NA=0.8), blue light, and a lab built heating stage were used, while keeping the sample in a nitrogen gas-flow chamber.

Saturation magnetization and anisotropy field (shown in Supplementary Note 1) were obtained by in-plane and out-of-plane magnetometry measurements.

# Results and discussion

## Structural characterization

Figure 1 (a) shows the X-ray diffraction pattern of samples A and B. The (002) peak of the single-crystal MgO substrate is visible at 42.9°. Further visible are the (002) and (004) reflections of MRG at 29.7° and 61.5°, respectively. The small, sharp reflections at 33.0°, 64.5°, and 77.4° originate from the sample holder. The films grow epitaxially on the MgO substrate with the relation MgO[001](100)||MRG[001](110), which was confirmed by reciprocal space maps (Figure 1(b)). From the peak positions of the (004)$_{MRG}$ reflections, we obtain c-parameters of 6.04 Å and 6.01 Å, for samples A and B, respectively. The epitaxial strain of about 1% induced by the MgO substrate ($\sqrt{2}\, a_{MgO} = 5.96$ Å) causes a tetragonal distortion of the MRG lattice, with elongation in out-of-plane direction. The tetragonal distortion is the reason for the perpendicular magnetic anisotropy of the epitaxial thin films. [13] The in-plane lattice parameters have been determined by reciprocal space maps of the (206)$_{MRG}$ reflections to be 5.98 Å (sample A) and 5.99 Å (sample B), both larger than $\sqrt{2}\, a_{MgO}$ indicating lattice relaxation, which gets more pronounced with increasing deposition temperature. The reciprocal space map of the (206)$_{MRG}$ peak of sample A is shown in Figure 1 (b). The relaxation of epitaxial strain is accompanied by misfit dislocations and grain boundaries, which disturb the coherence of the lattice. To confirm this, we determined the in-plane and out-of-plane X-ray coherence lengths $l_x$ and $l_z$ by fitting cuts through the (206) peaks in $q_x$ and $q_z$ directions. The results are shown in Table 1. Both in-plane and out-of-plane coherence lengths are larger for sample A than for sample B.

Table 1 Structural parameters of samples A and B

| Sample | $T_{dep}$ °C | c (Å) | a (Å) | c/a−1 (%) | $l_x$ (nm) | $l_z$ (nm) |
|---|---|---|---|---|---|---|
| A | 300 | 6.04 | 5.98 | 1.0 | 16.5 | 25.0 |
| B | 320 | 6.01 | 5.99 | 0.3 | 11.9 | 13.8 |

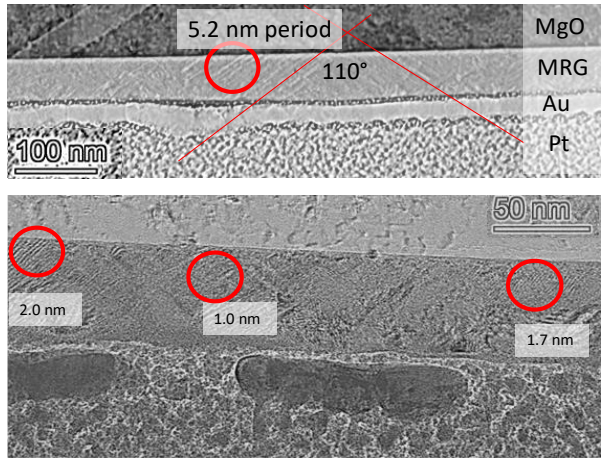

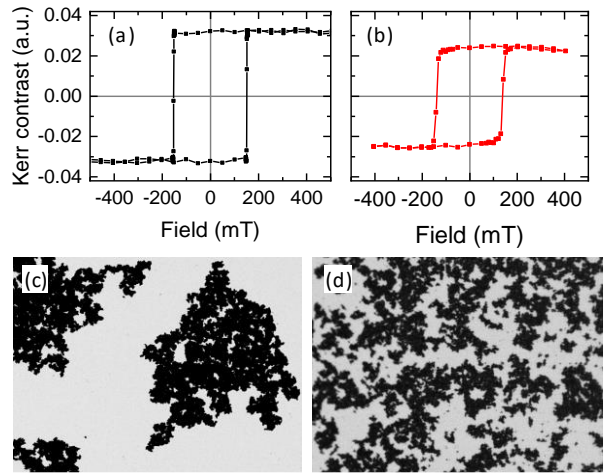

Figure 2 TEM cross sections of sample B. The spacings, and angles obtained from observed stripe contrast (marked in red circles and lines, respectively, fit to the expected values the {112} lattice planes.

For further structural analysis, TEM cross sections of sample B have been prepared in the $(110)_{MRG}$ orientation and images are shown in Figure 2. The visible stripe contrast in the red circle shows periodicity of multiples of 1.0 nm and angles with respect to the surface normal of around 55°. Therefore, the contrast originates from distorted {112} lattice planes. Structural defects, leading to these distortions may include twin boundaries, and misfit dislocations due to relaxation of the MRG lattice grown on the smaller MgO lattice.

Figure 3 MOKE hysteresis loops and domains during magnetization reversal for sample A (a) and (c) and B (b) and (d). The applied fields to obtain the domain patterns were in (c) −151 mT and (d) −139 mT. Fields of view are (c) 370 µm and (d) 74 µm.

## Magnetic domains

Figure 3 shows the magnetic hysteresis loops, obtained by Kerr microscopy for samples A and B. Both samples exhibit similar coercive fields of 151 mT for sample A and 139 mT for sample B and near perfect squareness. The magnetization reversal is very sharp in sample A and occurs within 3 mT whereas it takes around 30 mT for sample B. The visible negative slope in the saturation region of sample B is an artefact of the polar Faraday effect of the polarization microscope [12]. Shown in Figure 3(c) and (d) are corresponding domain images taken during the magnetization reversal at negative coercivity after saturation in +800 mT. The applied field is removed after 5 s to stop domain wall creep when taking the image. A few large domains of order 100 µm with irregular outlines are visible for sample A and many domains with sizes down to the submicron resolution limit of the microscope are seen for sample B.

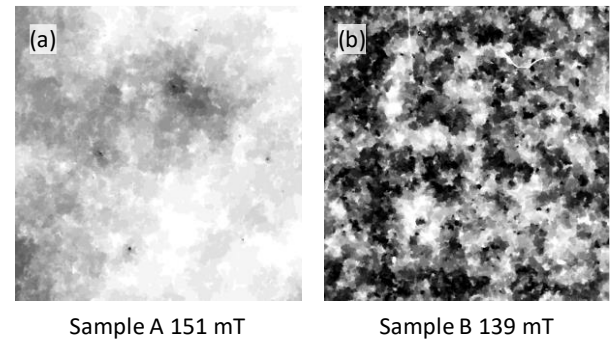

Sample A 151 mT    Sample B 139 mT

Figure 4 Average of several domain images taken under the same condition for (a) sample A averaging of 21 images taken at -151 mT (320 µm field of view) (b) sample B averaging of 23 images taken at -139 mT (64 µm field of view).

To check if the domain formation was deterministic, we repeated this process and averaged over several images taken in the same way. The results are shown in Figure 4. All images were binarized before averaging, so any white areas remain unchanged in all repetitions and black areas reverse every time. For sample A, a few black or dark grey spots can be identified, which indicate soft centres where the domains nucleate. The domain growth, however, is not really deterministic, as seen by the varying grey tones across the image. For sample B, nucleation sites are much denser and form visible patterns across the image with features parallel to the image borders. These directions correspond to [110] directions of the MRG lattice and is the same direction where strain contrast was seen in TEM images (cf. Figure 2) and corresponds to the easier in-plane magnetic anisotropy term.[14]

For quantitative analysis of the magnetization reversal, we studied the time evolution of magnetic domains at constant negative field after saturation in a large positive field. The results are shown in Figure 5 for sample A and in Figure 6 for sample B. For sample A the magnetic reversal occurs by dendritic growth of a few distinct, irregularly shaped domains. Remarkably, the rate of domain growth increases substantially when increasing the field by one millitesla, from taking 20 s reverse around half the sample's magnetization at 148 mT to 10 s for 148.5 mT and less than 6 s for 149 mT. Some areas in the panels of Figure 6 show parts of the domains in a lighter grey scale, most pronounced in the second panel of the last row. These are reversed *during* image acquisition, which for one image involves averaging over four exposures of 40 ms each.

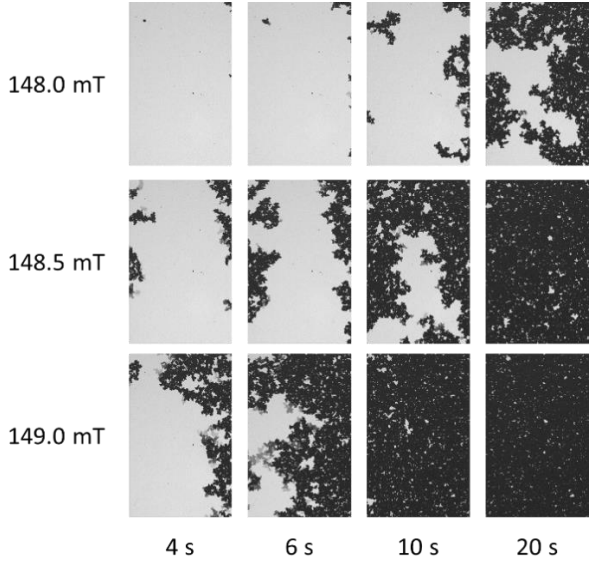

Figure 5 Magnetic aftereffect of sample A. Width of imaged area is 350 μm.

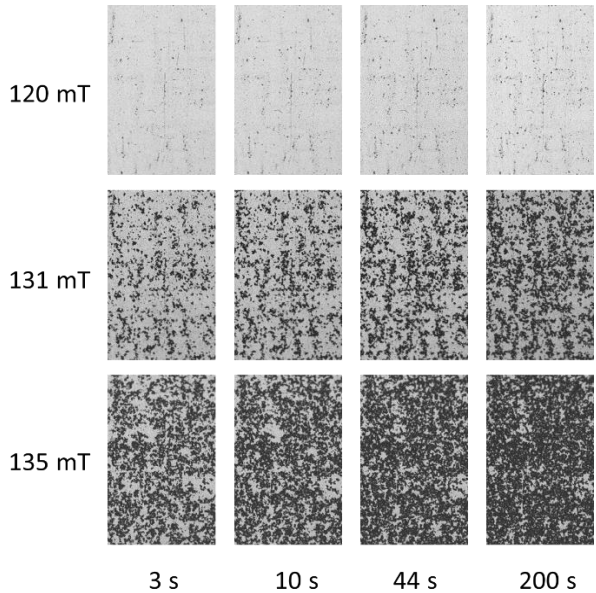

Figure 6 Magnetic aftereffect of sample B. Width of imaged area is 100 μm.

The few non-reversed (light) spots seen in the last panel at 149 mT did not switch within the timeframe of the measurement (140 s). They are hard centres and require higher field and several seconds to reverse. For sample B (Figure 6) magnetization reversal at constant field occurs by ongoing nucleation and growth of small domains. The timeframe for the magnetization reversal is clearly longer than for sample A and full magnetization reversal could not be observed for any field value because either the first recorded frame after field application is already partly reversed or the last frame after up to 800 s is not fully reversed.

The time dependent magnetization relaxation curves shown in Figure 7 (a) and (b) are based on Kerr contrast, but can be interpreted as $m(t) = \frac{M(t)+M(0)}{2\,M(0)}$. These data are used below to analyse activation volumes and energies. The increase of intensity for times under 1 s is related to the polar Faraday effect during the ramping of the magnetic field with a time constant of around 0.4 s.

The shape of the magnetic relaxation curves differs fundamentally between the two samples, showing "S"-shaped curves in the $m$-ln($t$) plot with decreasing width for increasing fields for sample A and almost linear behaviour for sample B across most of the investigated time frame.

Such different domain behaviours are common for thin films with perpendicular anisotropy and were already described for Au/Co/Au [10,15] and rare earth-transition metal thin films [16]. The differences are based on different activation energy distributions causing reversal dominated either by domain wall propagation or by nucleation. The magnetization reversal characteristics can be analysed based on a model developed by Fatuzzo [17] for polarization relaxation in ferroelectric materials and first used by Labrune et al. for perpendicular GdTbFe films [11].

The characteristic time, for which half the samples' magnetization is reversed ($t_{50}$) follows a phenomenological relation $t_{50} = t_0 \exp(\alpha\mu_0(H - H_0))$ with the activation time $t_0$ (taken as $10^{-10}$ s), $H_0$ the sample's intrinsic coercivity (without thermal activation, $H_0$ is in general larger than $H_C$, which is the coercivity read off magnetic hysteresis loops), $H$ the applied field, and $\mu_0$ the vacuum permeability. By comparison with an Arrhenius Law $\tau = \tau_0 \exp((E_A - \mu_0 H M_S V^*)/k_B T)$ with time constant $\tau$, the activation time $\tau_0 = t_0$ (a factor ln 2 is ignored here) the activation energy $E_A$, Boltzmann constant $k_B$ and the measurement temperature $T$, the Barkhausen volume $V^*$ can be derived from $\alpha = M_S V^*/k_B T$. The $\alpha$ parameter can be read off linear fits to the $\ln(t_{50})$-$H$-plots in Figure 7(c) and (d). The outlying $t_{50}$ values below 8 s were excluded from the analysis as there is strong influence of field ramping for short times. We obtain values of $\alpha = 1.54$ mT$^{-1}$ for film A and $\alpha = 1.56$ mT$^{-1}$ for film B which are almost identical despite the different magnetization reversal behaviour. Using the saturation magnetization (see Supplementary Figure 1) of 75 kA/m and 82 kA/m for samples A and B, respectively we obtain activation volumes of around $8.5 \times 10^{-5}$ μm$^3$ and $7.9 \times 10^{-5}$ μm$^3$ and characteristic lengths of $l^* = \sqrt{V^*/d} = 40$ nm and 39 nm for sample A and B, respectively, with $d$ being the film thickness. This is of the same order of magnitude as the domain wall width $\delta = \pi\sqrt{A/K_u}$, which is 20 nm for sample A and 26 nm for sample B, based on the anisotropy constant $K_u$, and the exchange stiffness $A$ (see Supplementary Note 1).

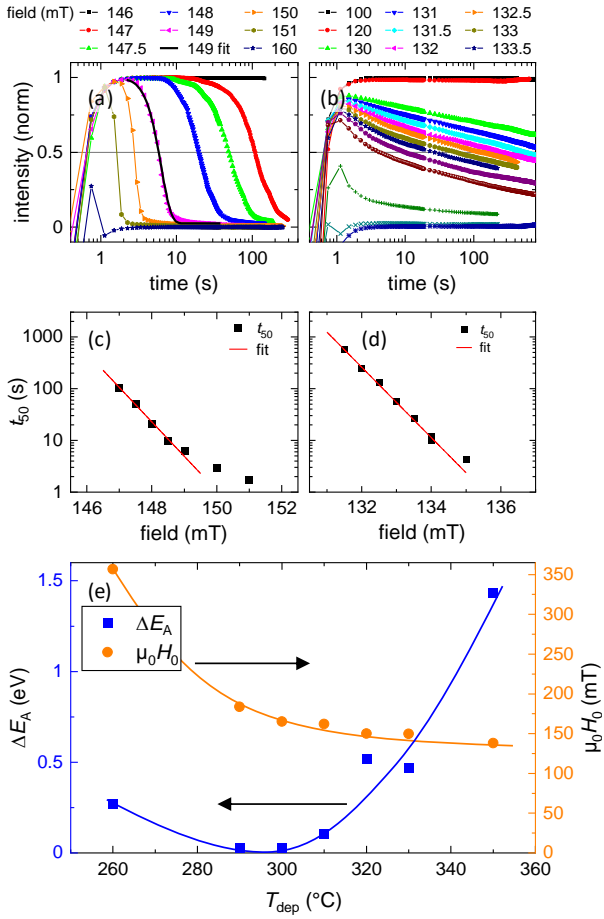

Figure 7 Magnetic aftereffect. Magnetic relaxation curves for different applied fields for (a) sample A and (b) sample B. The corresponding field dependence of $t_{50}$ is shown in (c) for sample A and (d) for sample B with fits to the linear regions on the logarithmic scale (red lines). The model yields for sample A: $\alpha = 1.54$ mT$^{-1}$ and $H_0 = 165$ mT and for sample B: $\alpha = 1.56$ mT$^{-1}$ and $H_0 = 150$ mT. (e) $\Delta E_A$ (blue squares, left axis) and $H_0$ (orange circles, right axis) as a function of the deposition temperature. The solid lines are visual guides. The minimum $\Delta E_A$ is reached for $T_{dep} = 290°C$ and the transition from domain wall motion dominated reversal to nucleation dominated reversal occurs around 0.2 eV < $\Delta E_A$ < 0.5 eV. $H_0$ decreases slightly with increasing $T_{dep}$.

The activation energy, which must be overcome for magnetization reversal by the action of the applied field and thermal activation, can also be derived from the Arrhenius law to be $E_A = k_B T(\alpha \mu_0 H_0)$ [11], giving 6.6 eV for sample A, and 6.1 eV for sample B. This corresponds to an energy density of ~12 kJ/m³ for both samples. Naturally, this is comparable but higher than the same determined from the coercive fields. However, this can only be understood as the centre of a distribution of activation energies. The width of this distribution can be inferred from the maximum slope of the $m(\ln(t))$-plots $\left(-\frac{dm(t)}{d\ln(t)}\right)_{max} = \frac{k_B T}{\Delta E_A}$, as was shown by Bruno et al. [18], which yields $\Delta E_A = 0.029$ eV for sample A and $\Delta E_A = 0.52$ eV for sample B. However, as for sample A, the magnetic reversal is governed by domain wall motion, which implies that an area of film can only switch if there is domain wall present in its vicinity. Therefore, the real distribution of activation energies, which are the domain wall propagation energies, is likely to be even narrower. The wider distribution of activation energies of sample B is supported by the linear behaviour at large times in the curves of Figure 7 (b), which corresponds to the widely used model for magnetic viscosity with $M(t) = M_0 - S \ln\left(1 + \frac{t}{t_0}\right)$ with viscosity coefficient $S$. The logarithmic, instead of exponential decay, results from the assumption of a wide range of activation energies, prohibiting the system from reaching thermal equilibrium even at long time scales. [19]

The original Fatuzzo-Labrune model states that
$m(t) = \exp\left[-2k^2\left(1 - (R_n t + k^{-1}) + \frac{1}{2}(R_n t + k^{-1})^2 - e^{-R_n T}(1 - k^{-1}) - \frac{1}{2}k^{-2}(1 - R_n t)\right)\right]$ (Eq. 1)

with $k = v/R$, domain wall velocity $v$, and nucleation rate $R_n$.[11] It is suitable for fitting the magnetic relaxation curves of sample A with $k = 30$ (black line in Figure 7 (a)), indicating domain wall motion dominated reversal. The curves of sample B cannot be fitted by Eq. 1, however, the overall shape of the data corresponds to $k = 0$, which is expected for a nucleation driven reversal.

The observation of increasing activation energy distributions and similar activation lengths was confirmed by analysing further films, deposited at $T_{dep}$ between 260°C and 350°C (see Figure 7(e)) showing that $\Delta E_A$ is minimum for $T_{dep} = 290°$ and increases strongly above $T_{dep} = 310°C$. The intrinsic coercivity $H_0$, decreases slightly increasing $T_{dep}$, however, only slightly above $T_{dep} = 290°$. Lower deposition temperatures lead to strongly increased coercivity and $\Delta E_A$. Activation energies range between 4.4 eV ($T_{dep} = 350°C$) and 9.0 eV ($T_{dep} = 260°C$) and characteristic lengths $l^*$ are between 32–40 nm. (See supplementary note 2)

## Discussion

The different magnetization reversal behaviour in samples A and B can be understood as follows: The wider distribution of activation energies of sample B leads to an abundance of soft centres allowing domain nucleation, which are separated by harder regions, impeding continuous domain wall motion process and therefore leading to nucleation dominated reversal process.[20]

The main reason for this is the distribution of local anisotropies, which govern the domain wall energies and also the energy barriers in Stoner-Wohlfarth-like switching. In reports on Au/Co/Au this has been attributed either to thickness variations or roughness of the Co layer [18] due to atomic steps in the underlayer [21], or variations of crystallite sizes [22]. In contrast to Co, the perpendicular anisotropy in MRG is not governed by the surface anisotropy contribution, so roughness is assumed to play a minor role on the activation energy; it is found to be small with respect to the film thickness, and similar for both films (cf. Supplementary Figure 4).

However, due to the higher deposition temperature, sample B exhibits a greater degree of lattice relaxation, manifested in shorter X-Ray coherence lengths. The uniaxial anisotropy in this material is governed by the tetragonal distortion of the unit cell, but not only is the mean tetragonal distortion lower for sample B than sample A, the smaller structural coherence length suggests a wider distribution of lattice parameters, meaning a wider distribution of local anisotropy. This directly leads to the wider range of activation energies we obtained from the magnetic aftereffect analysis. Other dislocations can also occur during the growth process, where initially formed islands merge together as the film gets thicker. This is discussed in the Supplementary note 3 for a series of films of varying thickness.

One precondition for application of this material in domain wall motion-based devices will be an ability to nucleate domain walls reproducibly at well-defined positions and move them at a sharply-defined depinning field or current. For these purposes, it would be best to use films like sample A, which shows a narrow distribution of activation energies and, therefore, the desired magnetization reversal behaviour, especially when compared to sample B. This seems to be the result of a better crystalline quality, as quantified by the larger structural coherence length and a lower degree of lattice relaxation, which results in a narrow distribution of activation energies.

It is seen in Figure 3 and Figure 5 that even the large domains of sample A exhibit irregular, fractal-like outlines. This is due to the random nature of the thermal activation involved in the magnetization reversal. This can be quantified by the fractal dimension of the domain *outline*, for which we obtain $D_f = 1.26$ from the last panel in Figure 5 (the outline of the continuous bright non-reversed region in the centre) using the box counting method. This is the same as the value obtained for Au/Co/Au films, where it was shown that with increasing applied field, the fractal dimension of the domain outline decreases from 1.26 below $H_C$ to ~1.0 above $H_0$. For high fields, energy barriers are overcome by the action of the applied field alone, without the need for thermal activation, which leads to smoother domain outlines. [23]

There is also a visible lacunarity (non-reversed spots) within the large domains of sample A (Figure 5), which are not the result of thermal activation but are rather indicative of slight inhomogeneities of the coercive field [15,24] (i.e. activation energy), or activation volume [21]. These hard centres are the counterparts of the domain nucleation centres on the other side of the activation energy range and likely caused by extrinsic defects.

The characteristic length scales of the activation volume do not correspond to the visible domain areas reversing simultaneously as seen in Figure 5 in mid-grey shading, which are more than an order of magnitude larger. Similar observations have been made for Au/Co/Au films with perpendicular anisotropy [23] and ascribed to avalanche behaviour. Areas around the Barkhausen volume, once switched, will reverse immediately afterwards as long as they are not pinned elsewhere.

The fact that the activation volume does not correspond to the size of simultaneously switched regions means that the activation length should not be interpreted as the distance between pinning centres. Similarly, the structural coherence length as determined from XRD measurements, which is of the same order as the activation length, does not correspond to the average spacing between pinning centres, either. This makes it challenging to quantify and correlate the activation length with structural parameters. The structural coherence length as determined from XRD measurements is a measure of the mean distance between defects that disturb the coherence of the lattice, like misfit dislocations or grain boundaries, but unlike point defects. But not every disturbance of the lattice is a magnetic pinning centre that would be visible in the domain pattern. The smallest visible length scale in the domain pattern is the coherence length of the domain wall $l_{wc}$ i.e. the length scale on which the domain wall bends. Governing this length scale is the distribution of pinning sites, along with the domain wall tension that inhibits sharp bends of the walls. This length scale has been determined from radially averaged autocorrelation analysis of domain wall images using the image processing software ImageJ [25]. For this we use images of thermally demagnetized films in order to maximise the length of domain wall in the image (see Supplementary note 4). We obtain $l_{wc} = 145$ nm for sample A and 120 nm for sample B, which is of the same order as the length scale determined from the TEM images. This length scale fits well with the appearance of the stripe contrast in the TEM in Figure 2, which appears at distances in the region of 100–200 nm and therefore can be interpreted as the location of possible pinning centres. However, it should be noted that for these samples the length scales obtained are close to the Abbe limit of 175 nm in the optical configuration used.

In order to further investigate the distribution of pinning centres based on domain observations in epitaxial MRG films we now focus the discussion on a third film (sample C) with larger domain wall coherence length of $l_{wc} = 240$ nm, which simplifies the analysis via optical methods. Details on the structural and magnetic characterization of sample C are found in the Supplementary Information. Here we focus on virgin domain patterns after thermal demagnetization, and first analyse formation of the domain pattern during controlled cooldown from above $T_C$, at a rate of ~3 K/min. In the second step we repeatedly thermally demagnetize the film and take images at exactly the same position after each demagnetization in order to generate a probability map of the domain wall locations.

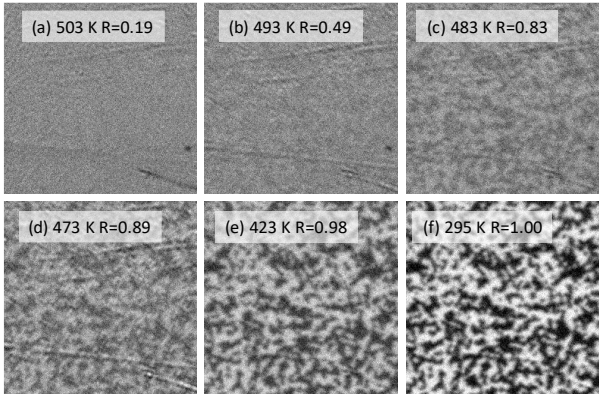

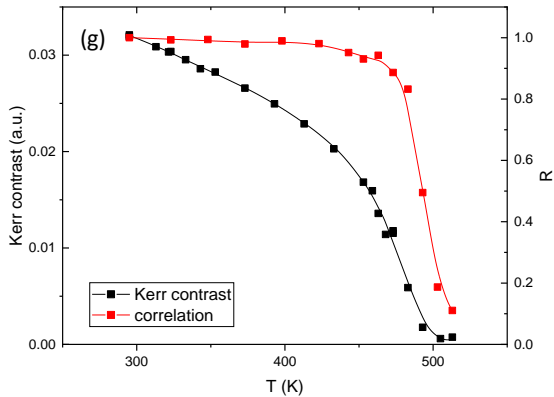

Figure 8 (a-f) Domain structure during cooldown from the Curie temperature. (g) Temperature dependence of Kerr contrast between light and dark domains (black curve, left scale) and the image correlation coefficient between virgin domain images taken at elevated temperature and at room temperature (red curve). The width of the images is 24 µm.

Figure 8 shows the domain structure during cooldown after thermal demagnetization. At high temperature (Figure 8(a)) no domain pattern is visible in the Kerr image. It emerges during cooldown (Fig. 8(b)) and is clearly visible below 483 K (Figure 8(b)-(f)). As the net magnetization as well as the anisotropy of the sample changes with temperature, it is expected that the domain pattern also changes to accommodate for the temperature dependence of domain wall energy and demagnetizing field. This however is *not* the case. The domain pattern seems to form just below $T_C$ = 500 K and stays *unchanged* down to room temperature. The behaviour is clear in Figure 8(g), where the Kerr contrast (left axis) and the correlation coefficient R (right axis) are plotted against temperature. The Kerr contrast increases gradually with decreasing temperature and follows the 4$c$ sublattice magnetization [26]. The correlation coefficient rises steeply from 0.19 to 0.83 between 503 K and 483 K which clearly indicates that the domain pattern at 483 K is essentially the same as the one at room temperature. Differences from a correlation coefficient of unity are due to the loss of contrast at high temperatures, which is only about a third of the room-temperature value.

Though it has to be noted that this sample has a compensation point at $T_{comp}$ = 420 K, just 80 K below $T_C$, where the saturation magnetization crosses zero (see Supplementary note 5). We estimate that the saturation magnetization is lower than $\mu_0 H_C$ for temperatures below 493 K, therefore, the influence of the demagnetizing field on the domain pattern is negligible. Nevertheless, the irregular nature and the observation that the domain pattern does not change with temperature shows that domain wall pinning is the dominant factor governing the domain pattern.

We now show how the domain pattern can be used to visualize prominent domain wall pinning sites by means of Kerr microscopy.

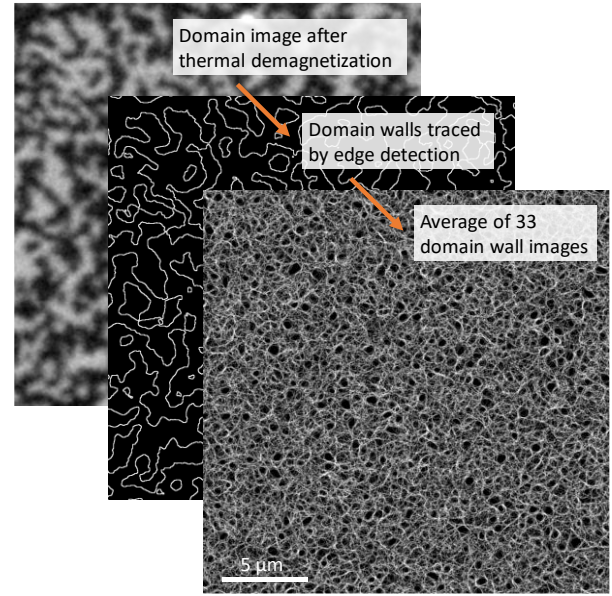

Figure 9 Visualization of pinning sites by domain imaging. Iimages are repeatedly taken after thermal demagnetization (background frame), the domain walls are traced by edge detection (center frame). Finally, an average of multiple (here 33) domain wall images from the same spot yields a "heat map" highlighting the domain wall pinning sites (foreground frame).

For this we take images after thermal demagnetization repeatedly at the same spot. This way, despite the domains (background frame in Figure 9) having random orientation - the correlation between subsequent images at the same spot was determined to be close to zero - domain walls (center frame) tend to be pinned at spots of low domain wall energy. An overlay of 33 of these domain wall images yields the foreground frame of Figure 9. The brightness at each spot maps the probability to find a domain wall at that pixel. It is seen that the pattern in the averaged domain wall image is not random, suggesting that domain wall tension alone is not the origin of the length scale of the domain pattern. It shows clear black areas that are avoided entirely by domain walls with no effective pinning sites and bright lines that are traced by domain walls in multiple images and therefore contain the pinning sites.

The size and distribution of the brightness minima can be used to quantify regions devoid of effective pinning centres. For simplicity, we focus our analyis on these black regions. Due to the continous nature of the domain walls, the pinning sites themselves are hidden in the nodes of the web of white lines on Figure 9, which are harder to analyse using image analysis software, but their length scale is of the same order. The majority of apparent pinning sites are surrounded by regions of low pinning.

The size of the black areas in Fig 9 (right panel) is of the order 300 nm and therefore corresponds to $l_{wc}$, but it is an order of magnitude larger than the structural coherence length and the length scales determined from the Barkhausen nucleation volume. Therefore, we were able to map magnetic pinning sites by means of Kerr microscopy and show that the 240 nm auto-correlation length scale of the domain walls is imposed by the average distance between prominent pinning centres.

This level of information on the structural defects present is otherwise only accessible via cross-sectional TEM as the pinning centres lie within the volume of the film.

Finally, it is important to discuss the implications of this work for devices which may depend on drivig magnetic domain walls by electrical current (such as spin-orbit torque-driven domain-wall logic gates). It has been shown that MRG exhibits exeptional damping-like spin-orbit torques [14], which cannot switch the magnetization direction of a magnetically saturated film, but should act substantially on the 4$c$ sublattice inside a Bloch wall. In a microwire geometry this should enable current induced-domain wall motion in a direction, which depends on the domain wall chirality. In order to explore this, further studies are needed that focus on the determination and control of domain wall chirality and domain wall velocities in thin strips of MRG that are optimized for minimum $\Delta E_A$. If patterned below a track-width of about 240 nm, devices with Hall-bar geometry, relying on the spontaneous (anomalous) Hall effect for detection, should exhibit switches which are abrupt down to timescales comparable with the inverse of the ferromagnetic resonance frequency, that is shorter than 10 ps.

## Summary and Conclusions

Our analysis of magnetic domain patterns during magnetization reversal for a set of MRG films with different deposition temperatures shows that that the magnetization reversal behavior changes radically from domain-wall-motion dominated behaviour to nucleation-dominated behaviour, within a very narrow range of deposition temperature, from 300°C to 320°C. By analyzing the magnetic aftereffect, we find that all investigated films show similar activation volumes with dimensions of 34–40 nm, and activation energies between 4.4 eV and 7.4 eV. Differences were mainly found in the distribution of activation energies that range between 28 meV and 1.4 eV. The samples with $\Delta E_A \geq$ 0.52 eV show reversal that is dominated by nucleation, while samples with $\Delta E_A \leq$ 0.10 eV show reversal dominated by domain wall motion. Comparing the structural parameters of the samples, we conclude that structural defects such as misfit dislocations are responsible for the range of activation energy since they locally disturb the anisotropy and act as pinning centres for domain walls.

Visualization of these pinning centres by a new Kerr imaging procedure, after thermal demagnetization, shows that the virgin domain pattern does not change appreciably during cooling the sample from $T_C$ to room temperature. It is shown that the prominent pinning sites are distanced by around the same length scale as the correlation length of the domain image, which is ~240 nm for the sample investigated.

In order to apply similar films in domain wall motion based spintronic devices, it is crucial that the pinning is of the type exhibited by sample A, or better yet by samples with smaller $\Delta E_A$. Large $\Delta E_A$ will negatively affect the distribution of depinning currents and the ability to nucleate domains in a controlled way. We have demonstrated that the deposition temperature is an effective handle on the optimization of the intensity of pinning in MRG. The estimate on the length-scale required for patterning, will serve as a guide for further studies, aimed towards spin-torque domain-wall motion-based devices, as well as ones aiming at achieving rapid (on the time scales below ~10 ps) switching of small volumes using spin-transfer or spin-orbit torques.

## Acknowledgements

This project has received funding from Science Foundation Ireland through contracts 16/IA/4534 ZEMS and 12/RC/2278 AMBER and from the European Union's FET-Open research programme under grant agreement No 737038. N. T. would like to acknowledge funding from the European Union's Horizon 2020 research and innovation programme under the Marie Skłodowska-Curie EDGE grant agreement No 713567. We also gratefully acknowledge funding from Northern Ireland's Department for Economy through US-Ireland grant USI 108.